\documentclass[a4paper,11pt]{article}
\usepackage{setspace}
\usepackage[left = 2cm, right=2cm, top=3cm, bottom=3cm]{geometry}
\usepackage[pdftex]{graphicx}
\usepackage{cite}
\usepackage{bm}
\usepackage{float}
\usepackage{amsmath,amssymb,latexsym}
\usepackage{color}
\usepackage[T1]{fontenc}
\usepackage{wasysym}
\usepackage{siunitx}
\usepackage{setspace}
\usepackage{relsize}
\usepackage{tikz}
\usepackage{lineno, blindtext}
\usepackage{authblk}
\usepackage[symbol]{footmisc}
\usepackage{lineno}
\usepackage[pdfencoding=auto, psdextra]{hyperref}

\begin{document}
	
	
	
	\renewcommand{\figurename}{Fig.}
	
	\title{\color{red}\textbf{Growth kinetics of interfacial patterns formed by the radial displacement of an aging viscoelastic suspension}}
	\author[1, $\dagger$]{Palak}
	\affil[1]{\textit{Soft Condensed Matter Group, Raman Research Institute, C. V. Raman Avenue, Sadashivanagar, Bangalore 560 080, INDIA}}
	\author[1, $\ddagger$]{Vaibhav Raj Singh Parmar}
	\author[1,*]{Ranjini Bandyopadhyay}
	
	\footnotetext[2]{palak@rri.res.in}
	\footnotetext[3]{vaibhav@rri.res.in}
	\footnotetext[1]{Corresponding Author: Ranjini Bandyopadhyay; Email: ranjini@rri.res.in}
	\maketitle
	\begin{abstract}
		When a soft glassy colloidal suspension is displaced by a Newtonian fluid in a radial Hele-Shaw geometry, the pattern morphology that develops at the interface is determined by the complex rheology of the former. We had reported in an earlier work~[Palak, V. R. S. Parmar, D. Saha and R. Bandyopadhyay, JCIS Open, 6 (2022) 100047] that a range of interfacial patterns can be formed by controlling the elasticity of the displaced suspension, the flow rate of the displacing fluid and the interfacial tension of the fluid pair. Interestingly, all the different  morphological features can be distinguished in terms of their areal ratios, defined as the ratio of the areas occupied by the fully-developed pattern and the smallest circle enclosing it. In a significant advance to this earlier work, we show here that a systematic study of spatio-temporal pattern growth can reveal important information about pattern selection mechanisms. We analyse the time-evolution of the patterns to reveal interesting correlations between their growth mechanisms and fully-developed morphologies. We believe that such systematic identification of the unique temporal features characterising pattern growth at the interface between an aging viscoelastic clay suspension and a Newtonian fluid can be useful in predicting and suppressing the onset and evolution of interfacial instabilities in the displacement of mud and cement slurries.

	\end{abstract}
	\definecolor{black}{rgb}{0.0, 0.0, 0.0}
	\definecolor{red(ryb)}{rgb}{1.0, 0.15, 0.07}
	\definecolor{darkred}{rgb}{0.55, 0.0, 0.0}
	\definecolor{blue(ryb)}{rgb}{0.01, 0.2, 1.0}
	\definecolor{darkcyan}{rgb}{0.0, 0.55, 0.55}
	\definecolor{navyblue}{rgb}{0.0, 0.0, 0.5}
	\definecolor{olivedrab(web)(olivedrab3)}{rgb}{0.42, 0.56, 0.14}
	\definecolor{darkraspberry}{rgb}{0.53, 0.15, 0.34}
	\definecolor{magenta}{rgb}{1.0, 0.0, 1.0}
	
	\newcommand{\blsquare}{\textcolor{black}{\small$\blacksquare$}}
	\newcommand{\hlsquare}{\textcolor{darkred}{\small$\square$}}
	\newcommand{\redtraingle}{\textcolor{magenta}{\small$\triangle$}}
	\newcommand{\oolive}{\textcolor{olivedrab(web)(olivedrab3)}{\large$\circ$}}
	\newcommand{\purpletraingle}{\textcolor{darkraspberry}{\small$\triangledown$}}
	
	\newcommand{\bltriangle}{\textcolor{black}{\small$\triangleup$}}
	\newcommand{\rcircle}{\textcolor{red(ryb)}{\large$\bullet$}}
	\newcommand{\rtraingle}{\textcolor{red(ryb)}{\small$\triangledown$}}
	\newcommand{\wine}{\textcolor{darkred}{\large$\bullet$}}
	\newcommand{\cyan}{\textcolor{darkcyan}{\large$\bullet$}}
	\newcommand{\owine}{\textcolor{darkred}{\large$\circ$}}
	\newcommand{\redcircle}{\textcolor{red}{\large$\circ$}}
	\newcommand{\ocyan}{\textcolor{darkcyan}{\large$\circ$}}
	\newcommand{\blue}{\textcolor{blue(ryb)}{\large$\bullet$}}
	\newcommand{\blbullet}{\textcolor{navyblue}{\large$\bullet$}}
	\newcommand{\olbullet}{\textcolor{olivedrab(web)(olivedrab3)}{\large$\bullet$}}
	\newcommand{\hollowblue}{\textcolor{blue(ryb)}{\large$\circ$}}
	\newcommand{\black}{\textcolor{black}{\large$\bullet$}}
	\newcommand{\hollowblack}{\textcolor{black}{\large$\circ$}}
	\newcommand{\bltria}{\textcolor{black}{\small$\triangle$}}
	\newcommand{\rtrai}{\textcolor{red(ryb)}{\large$\triangledown$}}
	
	\section{Introduction}
	Complex patterns spanning different length scales, such as ripples in sand, dendritic structures in snowflakes, the branching morphogenesis in vertebrate airways, and even the branching of trees and river networks, have been subjected to intense scientific scrutiny~\cite{CP}. Interfacial pattern formation at an unstable interface has been studied extensively in the context of many industrial applications, for example, petroleum extraction~\cite{orr1984use,gorell1983theory}, material processing and carbon sequestration~\cite{carbon}. There exists a wealth of literature on the growth and development of these instabilities that are often driven by fluid displacements~\cite{kondic1998non,park1994viscous,PALAK2021127405,lemaire1991viscous,ozturk2020flow,BALL2021104492,OSEIBONSU2016288} and are commonly studied in a quasi-two-dimensional Hele-Shaw (HS) cell comprising two transparent glass plates separated by an infinitesimal gap~\cite{PINILLA2021e07614}. Flow in the HS cell is governed by the balance between an external driving force and the viscous resistance of the displaced fluid, and is described by Darcy's law~\cite{Darcy}. The development of interfacial pattern morphologies can be tuned by introducing flow anisotropies, by either modifying the Hele-Shaw cell~\cite{etchedHS,groovedHS} or by controlling the bulk properties of the displaced viscoelastic fluid~\cite{PALAK2021127405,PALAK2022100047,azaiez2002stability,ozturk2020flow,lemaire1991viscous}. The complex flow behaviours of the displacing and displaced fluids significantly affect pattern morphologies, leading to characteristic features such as dendrites with tip-splitting and sides branching, jaggedness and fractures~\cite{PINILLA2021e07614,park1994viscous,kondic1998non}.

	Non-Newtonian viscoelastic fluids exhibit nonlinear rheological responses such as shear-thinning, shear-thickening, and non-zero yield stresses. The mechanical properties of viscoelastic fluids can have a significant effect on the growth kinetics of interfacial patterns generated during their displacement. It has been shown that repeated splitting of the evolving fingers can give rise to ramified structures with dense branching morphologies due to diffusion-controlled interfacial growth~\cite{Jacob}. Evolution of patterns with suppressed tip-splitting can be accompanied by an enhancement in the side shedding of branches~\cite{dendrites}. The displacement of carboxymethyl cellulose (CMC) solutions by liquid paraffin oil results in a transition from a tip-splitting interfacial pattern to a dendritic pattern as the concentration of the CMC solution is increased~\cite{CMC}. It was reported in this work that shear-thinning effects are maximal only at the early stages of the displacement process. The observed suppression of instabilities with increase in CMC concentrations was understood by considering the reduction in interfacial tension and the accompanying increase in viscous forces. The displacement of a shear-thinning fluid is driven by a competition between the increased velocity at the tip of the finger and simultaneous finger broadening due to large driving pressures~\cite{kondic1998non}. A study of the displacement of polymeric hydroxypropyl methyl cellulose (HPMC) solutions by air reported a deviation in the finger-tip velocity from the prediction of Darcy's law due to enhanced solution shear-thinning~\cite{HPMC}. We have previously proposed that increasing the viscosity ratio of the displacing and displaced fluids and the elasticity of the latter are efficient means of controlling instabilities during displacement of a dense shear-thinning granular cornstarch suspension by a glycerol-water mixture~\cite{PALAK2021127405}. Despite the low interfacial tension between the two fluids, the patterns displayed limited branching and features such as finger coalescence and proportionate growth during their temporal evolution. Previous studies reported that if the stored elastic energy in the displaced viscoelastic fluids exceeds the fracture energy, then viscoelastic fractures with pointed tips emerge and propagate faster than dendritic patterns~\cite{PALAK2022100047,lemaire1991viscous,hirata1998fracturing,ozturk2020flow}. The growth and morphologies of interfacial patterns in Hele-Shaw cell geometries are therefore extremely sensitive to the bulk properties of the displaced complex fluid. 
	
	\begin{figure}[!b]
		\includegraphics[width= 5.5in]{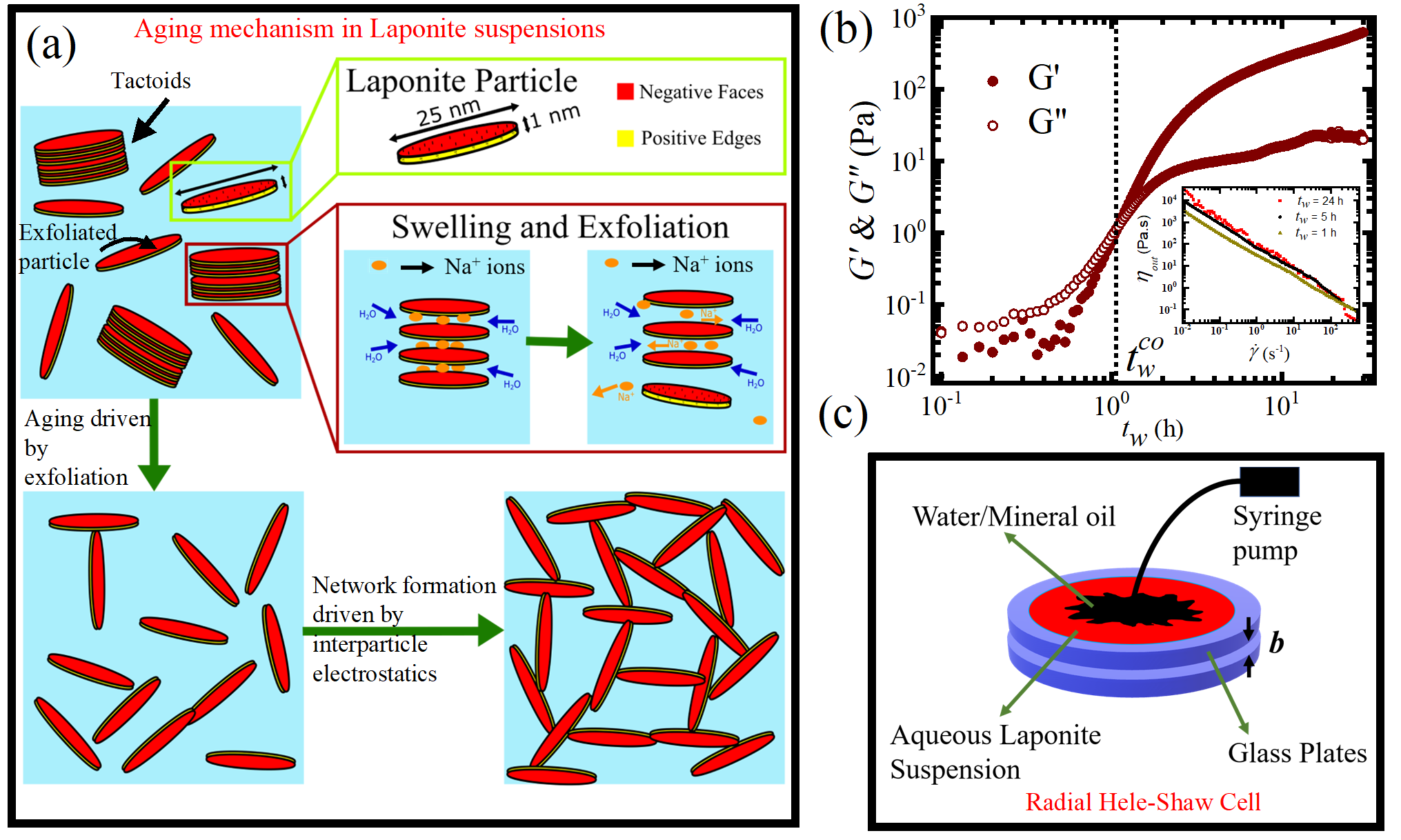}
		\centering
		\caption{\label{HS}\textbf{Aging and rheology of Laponite suspensions, and the experimental setup} \textbf{(a)} Schematic illustration of the aging dynamics of an aqueous suspension of Laponite particles. \textbf{(b)} Evolution of elastic modulus $G^{\prime}$ (\wine) and viscous modulus $G^{\prime \prime}$ (\owine) for a 3.25\% w/v Laponite suspension with increasing aging time $t_w$~\cite{PALAK2022100047}. Inset shows shear-thinning of 3.25\% w/v Laponite suspensions of various ages, $t_w$, wherein suspension viscosity decreases with increasing shear rate. \textbf{(c)} Schematic diagram of a radial Hele-Shaw cell. }
	\end{figure} 
	
	The displacement of clay suspensions plays a vital role in many fields, such as in the formation of river networks~\cite{roy1996patterns} and in drilling applications~\cite{drilling,cement}. The synthetic clay Laponite\textsuperscript{\textregistered} comprises approximately size-monodisperse disk-shaped particles of diameters 25-30 nm and thickness $\approx$ 1 nm (Fig.~\ref{HS}(a)). When clay particle tactoids (one-dimensional stacks) are dispersed in an aqueous medium, $\mathrm{Na^{+}}$ ions residing in the interlayer galleries of the tactoids start leaching out into the bulk due to osmotic pressure differences, thereby rendering a net negative charge on the faces of Laponite particles~\cite{van1977introduction,lagaly2013handbook}. The charge on the rims of the Laponite particles depends on medium pH, with the rims bearing positive charges below pH 11 due to hydration of magnesia groups~\cite{pH}. Aqueous suspensions of Laponite exhibit physical aging due to the gradual exfoliation of clay particles from the tactoids and the enhanced participation of $\mathrm{Na^{+}}$ ions in the Debye layer evolving around each clay particle. The heterogeneously charged Laponite particles therefore gradually self-assemble to form fragile microstructures composed of overlapping coins and house of cards (HoC) aggregates~\cite{delhorme2012monte}. These microstructures eventually percolate to form sample-spanning fragile networks. The aging process results in a spontaneous and gradual enhancement of sample rigidity~\cite{ali2016effect}, which we parameterise in terms of an aging time $t_w$, defined as the time elapsed since preparation of the sample. In this scheme, $t_w$ = 0 signifies the completion of the sample loading procedure and the onset of the aging process. By employing rheological measurements, we see that the elastic modulus (G$^\prime$) of the Laponite suspension exceeds the viscous modulus (G$^{\prime\prime}$) after an aging time $t_w^{co}$, signaling the onset of viscoelastic solid-like behaviour (Fig.~\ref{HS}(b)). At the lower aging times, G$^{\prime\prime}$ > G$^\prime$, indicating liquid-like rheology. The inset of Fig.~\ref{HS}(b) shows the shear-thinning rheology of Laponite suspensions of three different ages upon the application of shear strain rates.

	We had previously proposed that the aging time of the displaced colloidal clay suspension, flow rate of the displacing Newtonian fluid and miscibility of the fluid pair are reliable control parameters in the prediction of the global features of the observed interfacial patterns~\cite{PALAK2022100047}. We showed that the areal ratio, which we defined as the ratio of the areas of the interfacial pattern and the smallest circle enclosing it, can successfully predict the fully-developed interfacial pattern morphology. In addition to the striking structural features of fully-developed patterns, the growth of interfaces is driven by distinct mechanisms that need to be studied systematically~\cite{granular,complexstructure}. The objective of the present work is to characterise the growth kinetics of the distinct morphologies observed at the interface between a Newtonian fluid (displacing fluid) and an aqueous soft glassy colloidal suspension of Laponite clay (displaced suspension) in a radial Hele-Shaw geometry. Our experiments reveal the significant role of interfacial dynamics in determining the distinct morphologies of the fully-developed patterns. The shear and time-dependent rheology of the displaced soft glassy Laponite clay suspension and the wetting property of the displacing Newtonian fluid drive the formation of interfacial patterns with complex spatio-temporal features such as interacting fingers, side branch shedding, fracture propagation, development of interfaces characterised by flower petal-like and jagged shapes, etc. Interestingly, each of the above-mentioned morphologies grows differently. We show here that all the observed morphologies can be uniquely characterised by the time-dependent growth of their areal ratios and the velocities of the tips of the longest fingers. Such predictability of pattern morphologies based on their growth kinetics provides novel opportunities for understanding the formation of complex structures such as river networks~\cite{roy1996patterns}, dendrites in rechargeable batteries~\cite{battery}and spherulites~\cite{spherulites}.

	\section{\label{em}Experimental methods}
	The experimental setup used here to study interfacial pattern morphologies is a quasi-two-dimensional Hele-Shaw cell (Fig.~\ref{HS}(c)) that consists of two glass plates, each of radius 30~cm separated by a gap $b$ = $170~\mu m$. Teflon spacers were used to ensure a uniform gap between the plates. An aqueous suspension of Laponite of concentration 3.25\% w/v was prepared by adding dried Laponite XLG powder (BYK additives Inc.) in Milli-Q water (Millipore Corp., resistivity 18.2 M$\Omega$.cm). The Laponite suspension was first injected into the Hele-Shaw cell. After waiting for a time $t_w$ (referred to as the suspension aging time), the Newtonian fluid was introduced and the displacement of the suspension was recorded. We note here that $t_w$ = 0 corresponds to the time when the loading of the clay suspension in the HS cell was completed. Two different Newtonian fluids, water and mineral oil, were used to examine the miscible and immiscible displacements of Laponite suspensions in the Hele-Shaw cell. Both fluids were injected through a 3 mm hole on the top plate using a syringe pump (NE-8000, New Era Pump Systems, USA). The rheology (flow and deformation behaviour) of the displaced suspension was determined by the competition between suspension solidification due to aging and shear-thinning due to flow of the displacing fluid. To enhance the contrast at the interface, water was dyed with Rhodamine B ($\geq$95\% (HPLC)) and mineral oil was dyed with Oil red O procured from Sigma Aldrich. Interfacial pattern growth was recorded by a DSLR camera (D5200, Nikon, Japan) at a frame rate of 30 fps. The acquired images were converted to grayscale format and analysed using the MATLAB@2021 image processing toolbox. By following the protocols adopted in our previous work~\cite{PALAK2022100047}, RGB images were converted to binary form and areal ratios of all the patterns were estimated. The propagation velocity, $U$, of the interfacial pattern was estimated by tracking the growth of the longest finger. A stress-controlled Anton Paar MCR 501 rheometer was used for rheology measurements, in which freshly prepared Laponite suspensions were loaded in a concentric cylinder geometry and shear melted at a high shear rate of 500~$\mathrm{s^{-1}}$ to ensure a reproducible initial state. Next, the evolutions of the elastic and viscous moduli ($G^{\prime}$ and $G^{\prime\prime}$) of the Laponite suspension were recorded as a function of $t_w$ at an applied strain amplitude of 0.5\% and angular frequency 1 rad/s. A cone and plate geometry was used to measure flow curves of shear-thinning Laponite suspensions. More details about the experimental setups and the data analysis protocols can be found in our earlier publication~\cite{PALAK2022100047}.

	\section{\label{r&d}Results and Discussion}

	\begin{figure}[!t]
		\includegraphics[width= 5.0in]{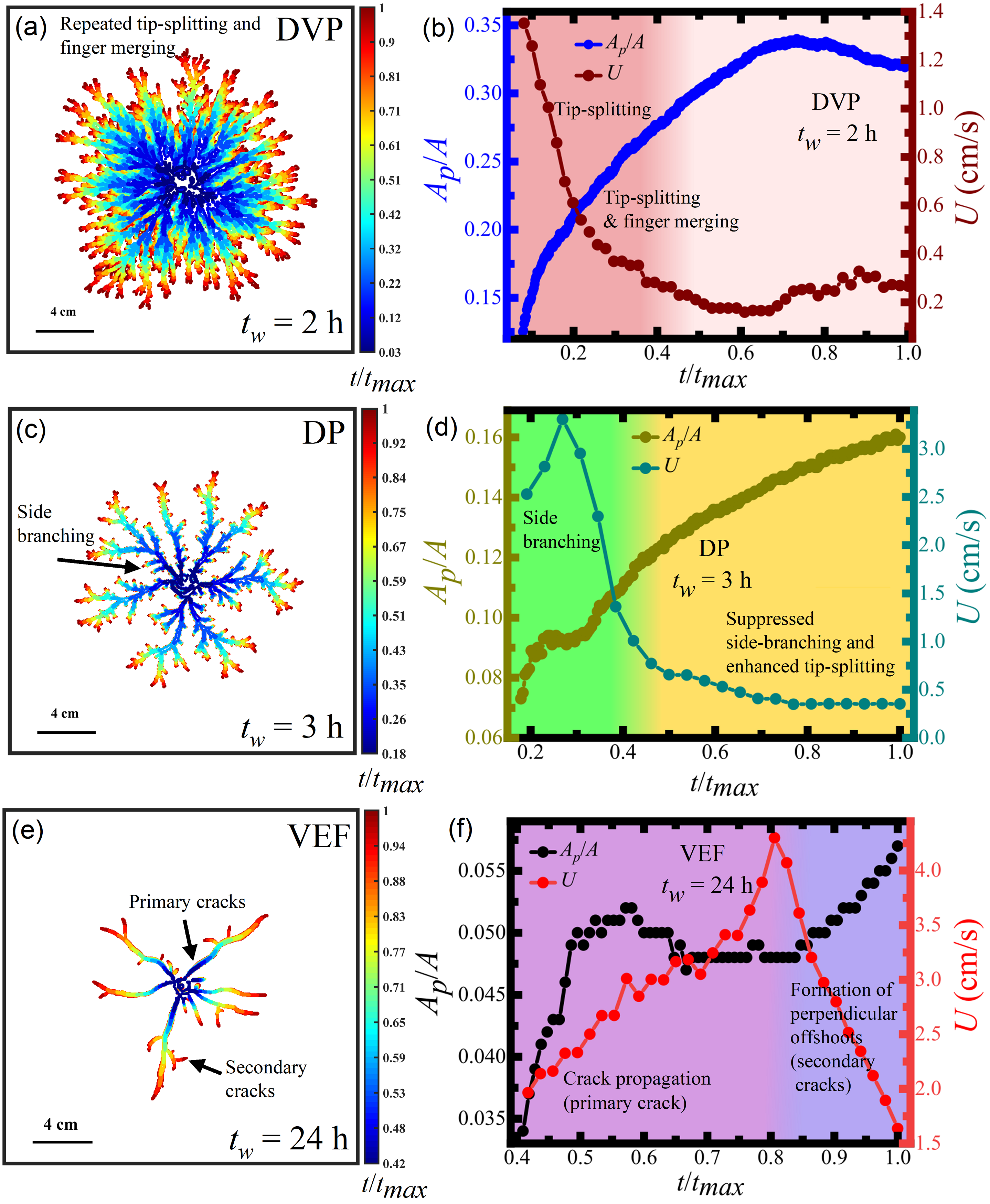}
		\centering
		\caption{\label{fpatterntw}\textbf{Temporal evolution of miscible displacement patterns at different suspension aging times.} The miscible (interfacial tension $\sigma$ = 0) displacement of 3.25\% w/v aqueous Laponite suspensions of different ages by water at a fixed flow rate $q$ = 5ml/min: \textbf{(a)} Dense viscous patterns (DVP) at $t_w$ = 2 h. The colours in the interfacial pattern in (a), defined by the colourbar on the right, have a one-to-one mapping with the pattern propagation time, normalised by the total experimental time $t_{max}$. \textbf{(b)} Areal ratio $A_p/A$ and instantaneous finger-tip velocity $U = dR/dt$ as a function of normalised time $t/t_{max}$ for the DVP displayed in (a). Here, $A_p$ is the area of the fully-developed pattern and $A = \pi R^2$ is the area of the smallest circle enclosing the entire pattern. Instantaneous velocity, $U$, is obtained by measuring the time-rate of propagation of the longest finger of length $R$. Transitions in the mechanisms driving the growth of the DVP in (a) are highlighted by different background colours in (b). \textbf{(c)} dendritic pattern (DP) at $t_w$ = 3 h. \textbf{(d)} $A_p/A$ and $U$ vs. $t/t_{max}$ for DP in (c). \textbf{(e)} Viscoelastic fractures (VEF) at $t_w$ = 24h. \textbf{(f)} $A_p/A$ and $U$ vs. $t/t_{max}$ for VEF in (e).}
	\end{figure}

	We investigate the growth of interfacial patterns when aqueous Laponite suspensions of concentration 3.25\% w/v and at predetermined aging times, 1 h $\leq t_w \leq$ 24 h, are displaced by miscible and immiscible fluids (water and mineral oil respectively) at various flow rates, $q$. Temporal evolutions of interfacial patterns when Laponite suspensions of various $t_w$ are displaced by miscible water (interfacial tension $\sigma$ = 0) are shown in Fig.~\ref{fpatterntw}. When a Laponite suspension of low age, $t_w$ = 2 h, is displaced by water, we observe that a very large number of fingers propagate radially outward with repeated tip-splitting at initial times, followed by frequent merging with neighbouring fingers at later times, as shown in Fig.~\ref{fpatterntw}(a). The resulting fully grown pattern morphology is identified as a dense viscous pattern (DVP). The colours in the interfacial patterns in Figs.~\ref{fpatterntw}(a,c,e) have a one-to-one mapping with the time of evolution of the pattern, normalised by the total experimental duration $t_{max}$, as indicated by the colourbars on the right of the figures. Since the Laponite suspension is liquid-like at low ages (Fig.~\ref{HS}(b)), the velocity at the base of the finger is large enough to ensure the effective displacement of the suspension, which results in the merging of neighbouring fingers. Due to such merging events during the formation of DVP, we observe many tiny irregularly-shaped regions wherein the displaced suspension is trapped. Consequently, a defining feature of DVP is the absence of well-defined interfacial boundaries. Several different interaction mechanisms between fingers, generated by the immiscible displacement of fluids having a significant viscosity contrast, were reported by Jackson $et$ $ al.$~\cite{coal1} $via$ numerical simulations. We see from our experiments that finger interactions are commonplace even for miscible displacement of a viscoelastic clay suspension at low sample ages. The left y-axis data in Fig.~\ref{fpatterntw}(b) quantifies the growth of interfacial patterns in terms of the development of their areal ratios $A_p/A$, defined as the area of the fully-developed pattern ($A_p$) normalised by the area of the smallest circle enclosing it ($A = \pi R^2$), and is estimated by following the protocols adopted in our previous work~\cite{PALAK2022100047}. The tip-splitting events and the frequent merging of neighbouring fingers cause an increase in $A_p/A$ with time, with $A_p/A$ reaching a constant value due to a balance between tip-splitting and finger merging at later times. We also track the propagation of the longest finger-tip of length $R$ and estimate its instantaneous velocity $U = dR/dt$ (Fig.~\ref{fpatterntw}(b), right y-axis). The plateauing of finger velocity $U$ at later times, as seen in Fig.~\ref{fpatterntw}(b), reflects steady pattern growth due to a dynamic equilibrium between finger-tip-splitting and merging events. The transition of the DVP growth kinetics from tip-splitting to a combination of side-branch shedding and tip-splitting, and eventually to steady growth at later times is highlighted by different colours in Fig.~\ref{fpatterntw}(b).

	The merging between neighbouring fingers is suppressed during the displacement of Laponite suspensions of intermediate ages ($t_w$ = 3 h, 5 h). We note the emergence of side-branching as a dominant propagation mechanism under these conditions. In contrast to DVP, the number of propagating fingers is far smaller in the dendritic pattern (DP) displayed in Fig.~\ref{fpatterntw}(c). DP growth is characterised by frequent side-branching at early times. This is in agreement with previous experiments on the displacement of shear-thinning carboxymethyl cellulose (CMC) solution by liquid paraffin oil. In this work, it was demonstrated that the local shear rates at the finger-tips achieved a constant magnitude after a certain time, such that shear-thinning of the displaced fluid affects pattern growth only during the early stages of pattern formation~\cite{CMC}. Furthermore, nonlinear simulations~\cite{kondic1998non} showed that local shear-thinning of the displaced fluid at the propagating finger-tips and the simultaneous broadening of the finger behind the tips due to the imposition of high driving pressures contribute to the formation of side branches. We note from Figs.~\ref{fpatterntw}(c-d) that shear-thinning-induced side-branching of the DP morphology is suppressed at later times and pattern formation is characterised by some tip-splitting events and eventually steady growth. Figure~\ref{fpatterntw}(d) shows an abrupt rise in $A_p/A$ at initial times, followed by a comparatively slower increase. This, in combination with the constant finger-tip velocity $U$ at later times during the growth of DP, indicates that side-branching is prominent only during the beginning of the experiment, with steady growth taking over thereafter. 
	
	On further increasing the aging time of the Laponite suspension to $t_w$ = 24 h, we note that the emergent elasticity of the suspension dominates the development of the growing interface during its displacement. When the driving pressure is large enough that the stored elastic energy is exceeded by the fracture energy, viscoelastic fractures (VEF, Fig.~\ref{fpatterntw}(e)) emerge. We observe that multiple tip-splitting is absent during the growth of VEF and the total number of fingers is minimal when compared to DVP and DP (Figs.~\ref{fpatterntw}(a-d)). VEF emerge with primary cracks at initial times, followed by the formation of secondary cracks at later times due to the release of large tensile stresses. The secondary cracks form approximately perpendicular to the primary cracks, which is a defining characteristic feature of a viscoelastic fracture~\cite{lemaire1991viscous}. Unlike DVP and DP, $A_p/A$ and $U$ for VEF show an initial sharp increase due to primary cracking (Fig.~\ref{fpatterntw}(e)). At later times, the occurrence of secondary cracks results in a second sharp increase in $A_p/A$, which is accompanied by a rapid decrease in $U$. Despite an identical flow rate of the displacing fluid in all the experiments discussed so far, VEF propagate faster in comparison to DVP and DP (Figs.~\ref{fpatterntw}(b,d)) and do not show steady growth at any stage of the experiment (Fig.~\ref{fpatterntw}(f)). $A_p/A$ and $U$ for miscible displacement experiments at several different aging times are shown in Supplementary Fig.~S1. These studies confirm that the morphological features and spatio-temporal growth regimes of the interfacial patterns formed by displacement of aging clay suspensions are robustly dependent on the aging time, $t_w$, of the suspension.
	
	\begin{figure}[!b]
		\includegraphics[width= 5.0in]{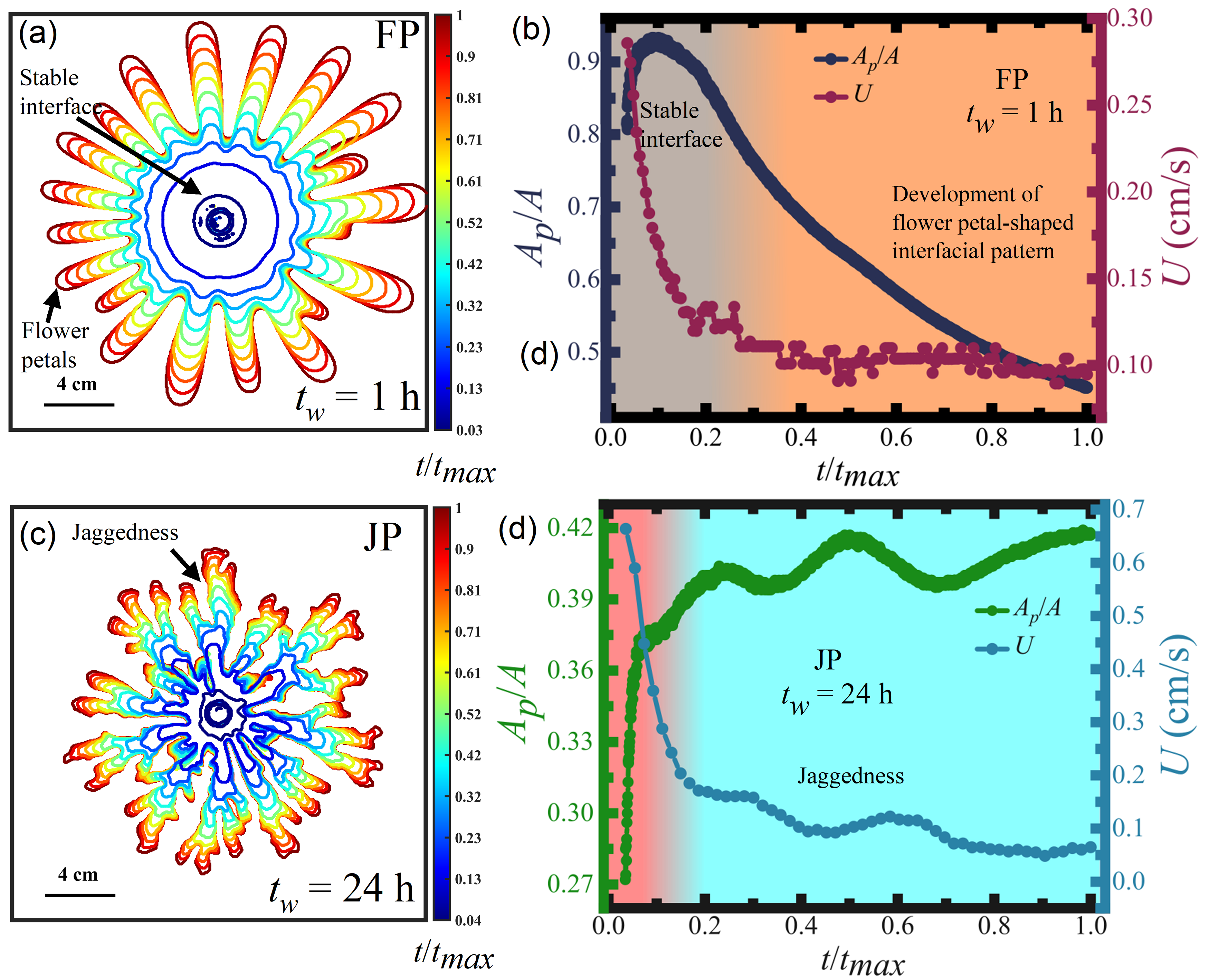}
		\centering
		\caption{\label{immiscibletw}\textbf{Temporal evolution of immiscible displacement patterns at various suspension aging times.} The immiscible (interfacial tension $\sigma \neq$ 0) displacement of 3.25\% w/v aqueous Laponite suspensions of different ages by mineral oil at a fixed flow rate $q$ = 5ml/min.  \textbf{(a)} Flower pattern (FP) at $t_w$ = 1 h. The colours in the interfacial pattern in (a), defined by the colourbar on the right, have a one-to-one mapping with the pattern propagation time, normalised by the total experimental time $t_{max}$. \textbf{(b)} $A_p/A$ and instantaneous velocity of the longest finger-tip, $U = dR/dt$, as a function of normalised time $t/t_{max}$ for FP in (a). Different colours in (b) indicate transitions in the growth profile of FP. \textbf{(c)} Jagged pattern (JP) at $t_w$ = 24 h. \textbf{(d)} $A_p/A$ and $U$ vs. $t/t_{max}$ for JP in (c).}
	\end{figure} 
	
	We next study the effects of interfacial tension on the immiscible displacement of aging Laponite suspensions by changing the displacing Newtonian fluid from miscible water to immiscible mineral oil ($\sigma$ $\neq$ 0). The inclusion of interfacial tension induces flow stabilisation by reducing the occurrence of tip-splitting events. At a low suspension aging time, $t_w$ = 1 h, we observe that the interface between the Laponite suspension and mineral oil is stable at an early stage of pattern growth, with interfacial perturbations appearing in the form of thick finger-like protrusions only at later times. Since these patterns, such as the one displayed in Fig.~\ref{immiscibletw}(a), bear a resemblance to flower petals, we call them flower patterns (FP). The stable circular interface at the initial stage of pattern growth results from a delay in onset of instability due to non-zero interfacial tension and resembles observations in previous experiments involving Newtonian fluid pairs~\cite{paterson1981radial,bischofberger2015island} and oil-clay suspension interfaces~\cite{PALAK2022100047}. We note from Fig.~\ref{immiscibletw}(b) that $A_p/A$ values for FP decrease with time due to the emergence of thick protrusions without tip-splitting. Interestingly, the steady growth of the longest finger-shaped perturbation is evident from the approximately constant values of $U$ during much of the experimental duration. In contrast, when we displace a highly aged Laponite suspension ($t_w$ = 24 h) with mineral oil, the evolving patterns are characterised by many more fingers than seen at earlier ages, while even displaying occasional tip-splitting events. The competition between shear-thinning and growth in elasticity of the displaced suspension at high $t_w$ results in jagged patterns (JP, Fig.~\ref{immiscibletw}(c)). We see from Fig.~\ref{immiscibletw}(c) that the fingers have unique wavy profiles along the azimuthal direction (shown by a black arrow), which gives rise to jaggedness in the morphology of the evolving pattern. In contrast to flower patterns, $A_p/A$ for the jagged pattern increases due to tip-splitting events during its growth. The instantaneous finger-tip velocity $U$ for JP decreases with time, but shows a striking oscillatory behaviour during the later stages of pattern evolution. We believe that the observed waviness in $U$ and $A_p/A$ for the JP arises due to disruption and subsequent reformation~\cite{Biswas} of the percolating fragile elastic microstructures in the highly aged displaced clay suspension. Such interfacial jaggedness was first observed during the displacement of foam, an yield stress fluid, by immiscible air, and was believed to arise from the elastic rearrangements of bubbles~\cite{park1994viscous}. $A_p/A$ and $U$ for immiscible patterns, obtained by displacing aging aqueous Laponite suspension at several different aging times, are shown in Supplementary Fig.~S2. In summary, the morphologies and growth regimes of interfacial patterns can be completely transformed by tuning the mechanical properties of the displaced suspension and the interfacial tension of the fluid pair.

	\begin{figure}[!t]
		\includegraphics[width= 5.0in]{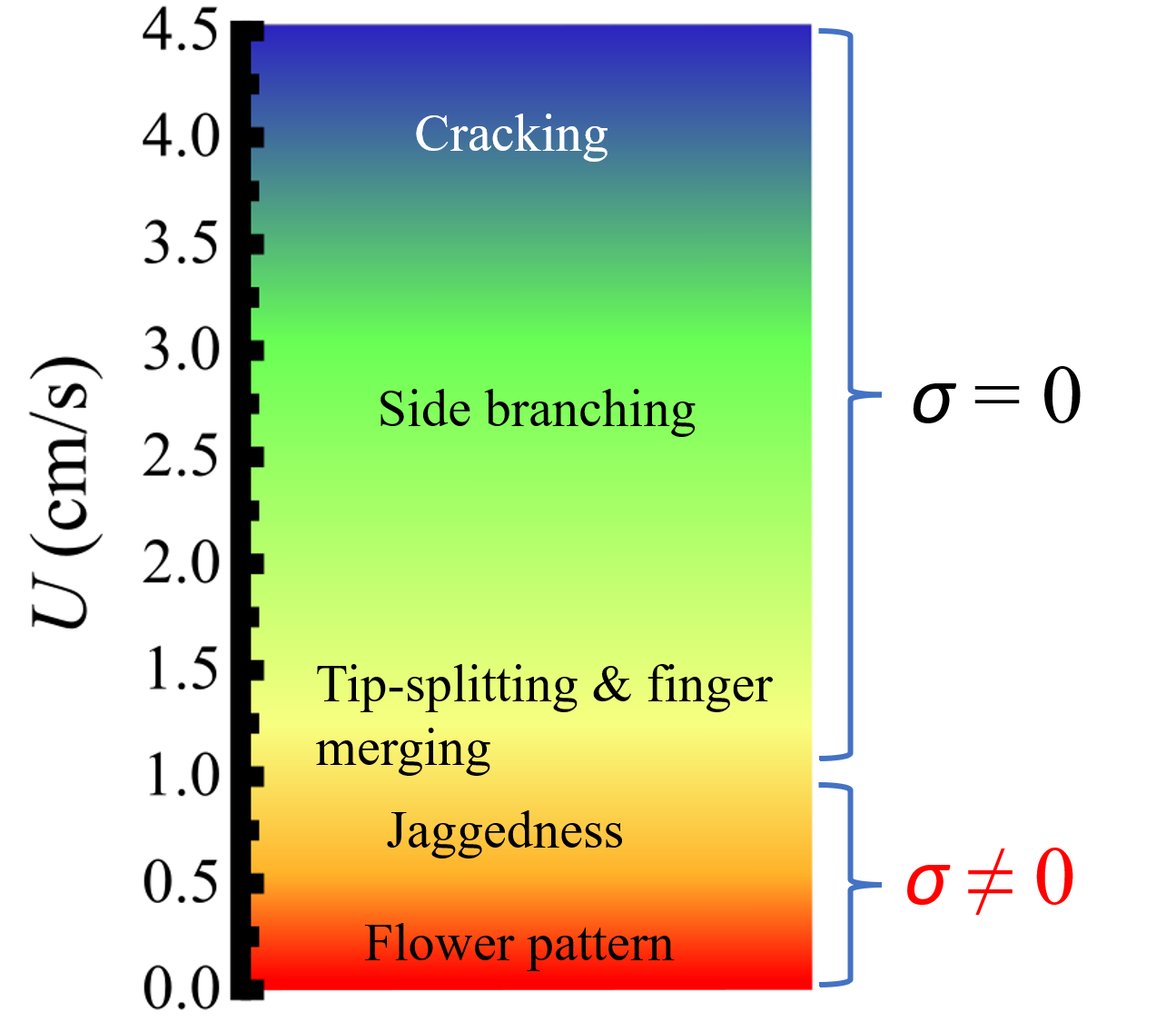}
		\centering
		\caption{\label{dRdt}\textbf{Distinct mechanisms of pattern evolution depend on the velocity of the longest finger-tip, $\boldsymbol{U}$.} The different interfacial pattern morphologies can be characterised and segregated by considering only the interfacial tension $\sigma$ and $U$.}
	\end{figure}
	
	Figure~\ref{dRdt} summarises and segregates the distinct growth mechanisms of the observed pattern morphologies in terms of $U$, the instantaneous velocity of the longest finger-tip. Interfacial tension stabilises the interface and results in patterns that propagate at very low velocities, while developing petal-like and, subsequently, jagged morphologies as the displaced Laponite suspension becomes more elastic with age (Fig.~\ref{immiscibletw}). In the absence of interfacial tension, the formation of sharp tips costs less energy. Depending on the elasticity, and therefore the aging time, $t_w$, of the displaced clay suspension, we note the growth of miscible patterns characterised by tip-splitting, frequent merging between fingers, side-branching and cracking. Despite the fact that the flow rate of the displacing fluid was kept constant in the above experiments, cracks are seen to propagate the fastest when compared to all other interfacial pattern morphologies. The flow properties of aging Laponite suspensions change dramatically when sheared, particularly at very high suspension aging times~\cite{PALAK2022100047,ruzicka2011fresh,misra}. We have therefore also studied the temporal evolutions of interfacial patterns when a highly aged Laponite suspension of $t_w$ = 24 h, is displaced by a Newtonian fluid at various flow rates, $q$. We had previously shown that the displacement of Laponite suspensions of progressively higher ages, $t_w$, and therefore of higher elasticities, $G^{\prime}$, can produce the same sequence of interfacial patterns that are obtained by decreasing the injection flow rate of the displacing fluid, $q$~\cite{PALAK2022100047}. The temporal evolutions of all the interfacial patterns at various flow rates, $q$, of the displacing Newtonian fluid are shown in Supplementary Fig.~S3. Variations in $A_p/A$ and $U$ of the displacement patterns (Supplementary Fig.~S4) resemble those seen in our earlier experiments at different suspension aging times $t_w$ (Figs.\ref{fpatterntw}-\ref{immiscibletw} and Supplementary Figs.~S1-S2), with pattern features obtained with low flow rates of the displacing fluid being very similar to those obtained by displacing a clay suspension of a higher age. Further details are provided in the supplementary material section ST1. Our experiments indicate that desired growth of pattern morphologies can be attained either by changing the aging time, $t_w$, of the Laponite suspensions, the flow rate, $q$, of the displacing Newtonian fluid, or the interfacial tension of the fluid pair. We show here, therefore, that besides controlling the morphological features of fully-developed patterns at the interface between a Newtonian fluid and an aging clay suspension~\cite{PALAK2022100047}, the aging time of the displaced suspension, $t_w$, the shear rate exerted by the displacing fluid, quantified in terms of its flow rate $q$, and the interfacial tension can also reliably predict the temporal evolution of interfacial instabilities.

	\section{\label{se:sac} Conclusions}
	Aging colloidal clay suspensions show spontaneous aging due to the formation of system-percolating fragile microstructures driven by the evolution of electrostatic interactions between the constituent particles~\cite{ali2016effect,delhorme2012monte,bandyopadhyay2004evolution}. In a previous contribution, we had reported that the evolving elasticity and shear-thinning rheology of soft glassy clay suspensions lead to the generation of a wide range of patterns when the aging suspension is displaced at various flow rates by Newtonian fluids of different miscibilities~\cite{PALAK2022100047}. We had reported that displacing a clay suspension of increasing age, $t_w$, at a fixed flow rate, $q$, generates the same sequence of patterns as seen during the displacement of the suspension at a fixed $t_w$ while lowering $q$ of the displacing fluid.

	In this work, we systematically study the growth kinetics of the distinct morphologies observed at the interface between a Newtonian fluid (displacing fluid) and an aqueous soft glassy colloidal suspension (displaced suspension). Dense viscous patterns (DVP) with repeated splitting of the evolving fingers and resembling diffusion-controlled interfacial growth~\cite{Jacob}, are observed for miscible displacements of clay suspensions of low ages. With increasing age of the displaced clay suspension, patterns with enhanced side shedding of branches are recorded and identified as dendritic patterns (DP)~\cite{dendrites,CMC,kondic1998non}. This is followed by emergence of viscoelastic fractures (VEF) for the highest suspension ages~\cite{lemaire1991viscous,hirata1998fracturing,ozturk2020flow}. We note the development of complex kinetic features during the growth of miscible interfacial patterns, such as repeated tip-splitting and merging of fingers in DVP, side-branching in DP, and crack propagation in VEF. We effectively quantify and distinguish the spatio-temporal growth of the distinct pattern morphologies in terms of their areal ratios and the instantaneous velocities of the tips of the longest propagating fingers. For displacement experiments at non-zero interfacial tension, we observe a delay in onset of instability and an initial stable region in the interfacial pattern when the age of the displaced Laponite suspension is low. This is followed by the emergence of thick protrusions, such that the instability takes the shape of flower petals. This result is reminiscent of earlier experiments involving Newtonian fluid pairs~\cite{paterson1981radial,bischofberger2015island}. Furthermore, we observe a jagged interfacial pattern when the elasticity of the displaced suspension is very high. Jaggedness at the interface was also previously observed during the immiscible displacement of an yield stress material (foam) by air~\cite{park1994viscous}. The evolution of the stable central region in flower patterns and finger-tip-splitting in jagged patterns are reflected in the temporal variations of the tip velocities of the longest fingers and the pattern areal ratios.
	
	To the best of our knowledge, this is the first study that characterises a wide variety of fully-developed pattern morphologies, generated using miscible and immiscible fluid pairs, in terms of pattern growth kinetics. By systematically measuring and analysing the evolutions of areal ratios and tip velocities of the longest propagating fingers, we show here that all the distinct interfacial pattern morphologies can be segregated on the basis of their temporal growth. The ability to distinguish interfacial patterns, whether through identification of their fully-developed morphologies, or by the range of growth mechanisms uncovered here, can provide novel opportunities to control the growth of unstable interfaces~\cite{CP}.
	
	To summarise, we show here that it is possible to predict fully-developed interfacial pattern morphologies by studying the different stages of pattern growth. An extension of this work could be a systematic study of interfacial instabilities obtained by displacing a variety of non-Newtonian materials with distinct underlying medium structures and under different physicochemical conditions. Interparticle interactions in charged colloidal clay suspensions, for example, can be altered by changing clay concentration and medium temperature, and by incorporating additives such as salts and acids~\cite{saha2015dynamic,Venketesh,Shahin2012}. Such interventions are known to alter the time-dependent rheology of soft glassy clay suspensions~\cite{misra,Venketesh}. Given the  importance of interfaces in fundamental fluid mechanics and in materials processing, for example in the displacement of mud and cement slurries, we believe strongly that our work will trigger further research on the onset and growth of complex interfacial patterns.

	\section*{Declaration of Competing Interest}
	The authors declare that they have no known competing financial interests or personal relationships that could have appeared to influence the work reported in this paper.
	
	\section*{Data Availability}
	Source data are available for this paper from the corresponding author upon reasonable request.
	
	\section*{Acknowledgments}	 
	We thank Raman Research Institute (RRI, India) for funding our research and Department of Science and Technology Science and Education Research Board (DST SERB, India) grant EMR/2016/006757 for partial support.
	
	\bibliographystyle{elsarticle-num}
	\bibliography{ref}

	\pagebreak
\renewcommand{\figurename}{Supplementary Fig.}
\begin{center}
	\textbf{\LARGE Supplementary Material}\\
	\vspace{0.4cm}
	\textbf{\LARGE \textcolor{red}{Growth kinetics of interfacial patterns formed by the radial displacement of an aging viscoelastic suspension}}\\
	\vspace{0.4cm}
	{Palak}$^\dagger$, {Vaibhav Raj Singh Parmar}$^\ddagger$, and {Ranjini Bandyopadhyay}$^*$\\
	\vspace{0.4cm}
	\textit{Soft Condensed Matter Group, Raman Research Institute, C. V. Raman Avenue, Sadashivanagar, Bangalore 560 080, INDIA}
	
	\date{\today}
\end{center}

\footnotetext[2]{palak@rri.res.in}
\footnotetext[3]{vaibhav@rri.res.in}
\footnotetext[4]{Debasish.Saha@uni-duesseldorf.de}
\footnotetext[1]{Corresponding Author: Ranjini Bandyopadhyay; Email: ranjini@rri.res.in}
\maketitle

\definecolor{black}{rgb}{0.0, 0.0, 0.0}
\definecolor{red(ryb)}{rgb}{1.0, 0.15, 0.07}
\definecolor{darkred}{rgb}{0.55, 0.0, 0.0}
\definecolor{blue(ryb)}{rgb}{0.01, 0.2, 1.0}
\definecolor{darkcyan}{rgb}{0.0, 0.55, 0.55}
\definecolor{navyblue}{rgb}{0.0, 0.0, 0.5}
\definecolor{olivedrab(web)(olivedrab3)}{rgb}{0.42, 0.56, 0.14}
\definecolor{darkraspberry}{rgb}{0.50, 0.0, 1.0}
\definecolor{magenta}{rgb}{1.0, 0.0, 1.0}
\newcommand{\pcircle}{\textcolor{darkraspberry}{\large$\bullet$}}
\newcommand{\phex}{\textcolor{darkraspberry}{\large$\varhexagonblack$}}
\newcommand{\rdhexagon}{\textcolor{red(ryb)}{\small$\blackdiamond$}}

\setcounter{table}{0}
\renewcommand{\thetable}{S\arabic{table}}%
\renewcommand{\tablename}{Supplementary Table}
\setcounter{figure}{0}
\makeatletter 
\renewcommand{\figurename}{Supplementary Fig.}
\setcounter{figure}{0}
\makeatletter 
\renewcommand{\thefigure}{S\arabic{figure}}
\setcounter{section}{0}
\renewcommand{\thesection}{ST\arabic{section}}
\setcounter{equation}{0}
\renewcommand{\theequation}{S\arabic{equation}}

	\begin{figure}[H]
	\includegraphics[width= 5.0in]{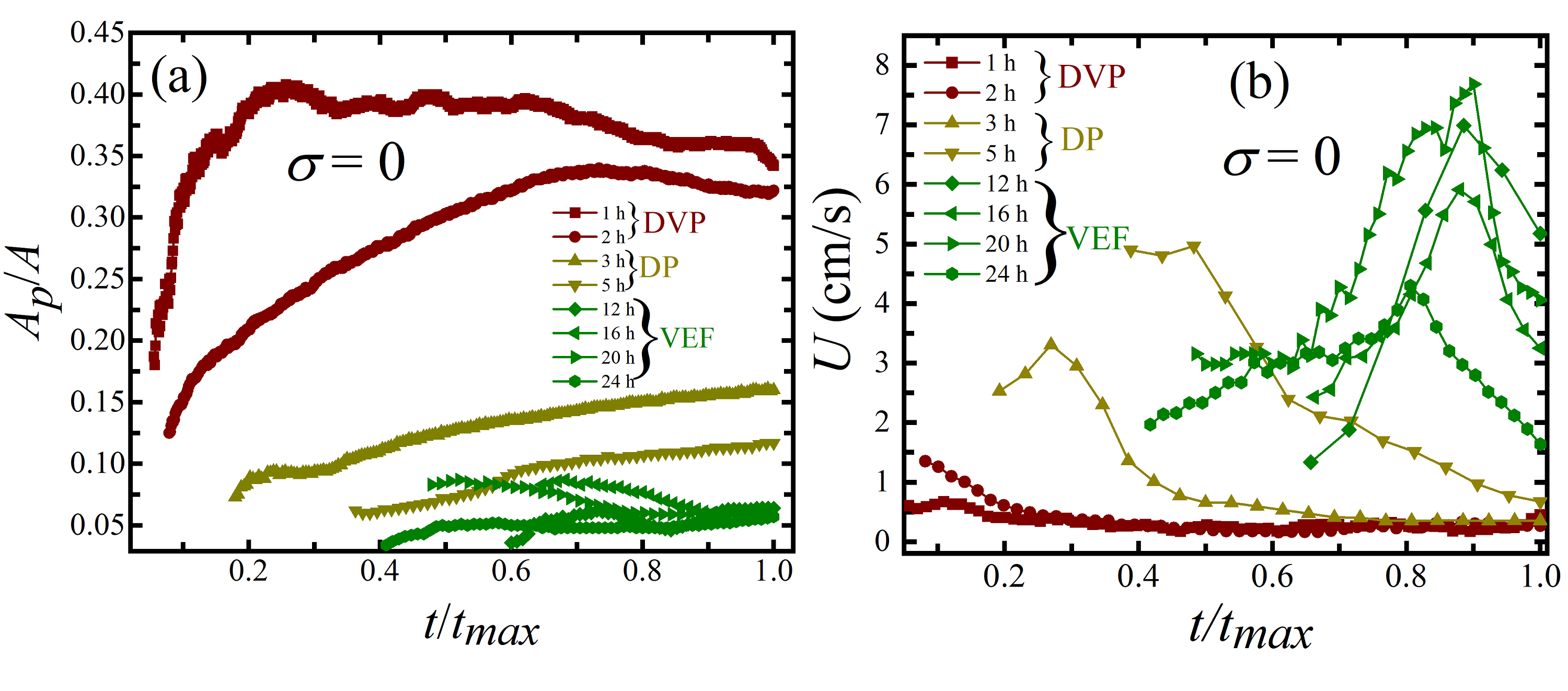}
	\centering
	\caption{\label{misctw}\textbf{Areal ratios and instantaneous velocities of miscible patterns with suspension aging time $\boldsymbol{t_w}$ as a control parameter.} \textbf{(a)} $A_p/A$ as a function of pattern propagation time $t$, normalised by the total experimental time $t_{max}$ for miscible (interfacial tension $\sigma$ = 0) patterns observed during the displacement of 3.25\% w/v aqueous Laponite suspensions of different ages by water at a fixed flow rate $q$ = 5ml/min. Here, $A_p$ is the area of the fully-developed pattern and $A = \pi R^2$ is the area of the smallest circle enclosing the entire pattern. \textbf{(b)} Instantaneous velocity $U$ vs. $t/t_{max}$ for different aging times $t_w$.}
\end{figure}

\begin{figure}[H]
	\includegraphics[width= 5.0in]{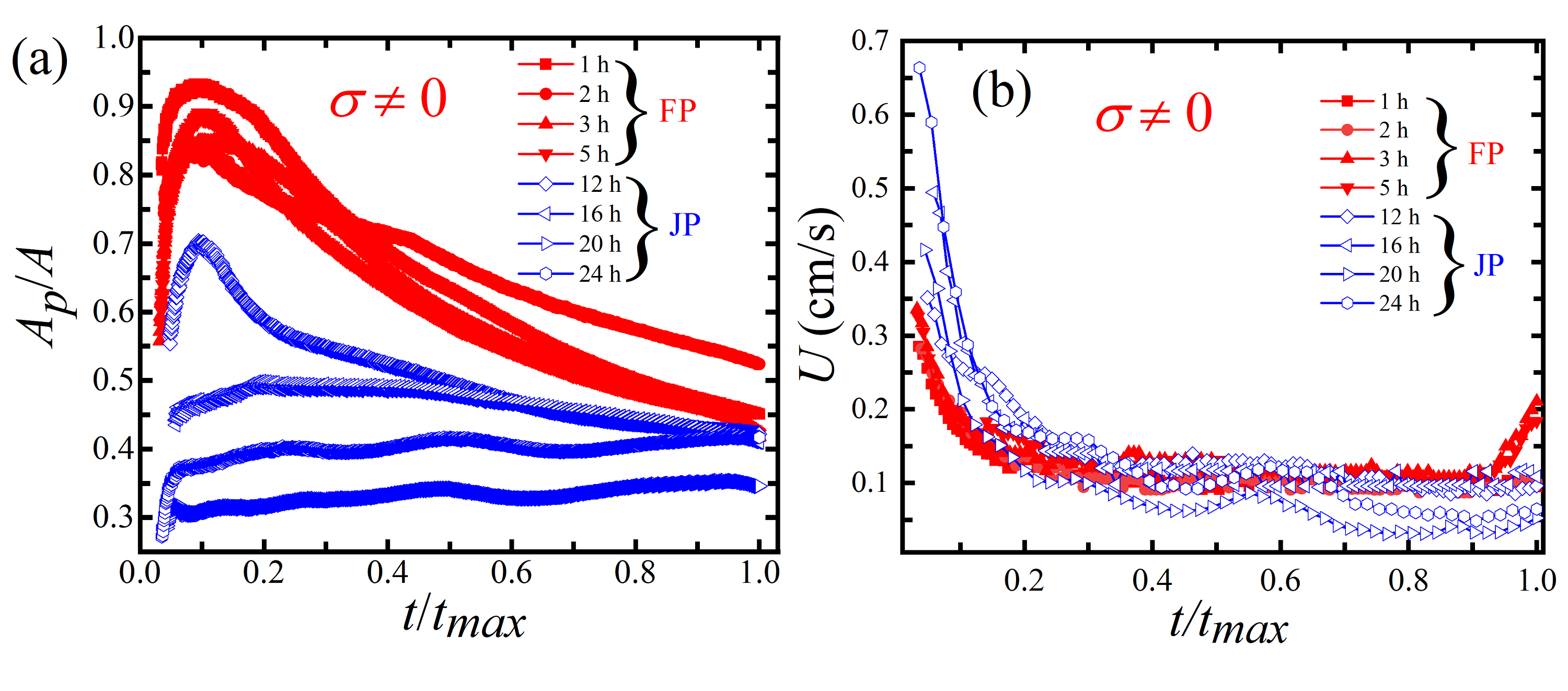}
	\centering
	\caption{\label{immtw}\textbf{Areal ratios and instantaneous velocities of immiscible patterns with suspension aging time $\boldsymbol{t_w}$ as a control parameter.} \textbf{(a)} $A_p/A$ as a function of pattern evolution time $t$, normalised by with total experimental time $t_{max}$, for immiscible (interfacial tension $\sigma$ = 0) patterns observed during the displacement of 3.25\% w/v aqueous Laponite suspensions of different ages by mineral oil at a fixed flow rate $q$ = 5ml/min. \textbf{(b)} Instantaneous velocity $U$ vs. $t/t_{max}$ for different aging times $t_w$.}
\end{figure}

\section{Displacement of Laponite suspensions at $t_w$ = 24 h by Newtonian fluids at various flow rates.}

\begin{figure}[!t]
	\includegraphics[width= 5.5in]{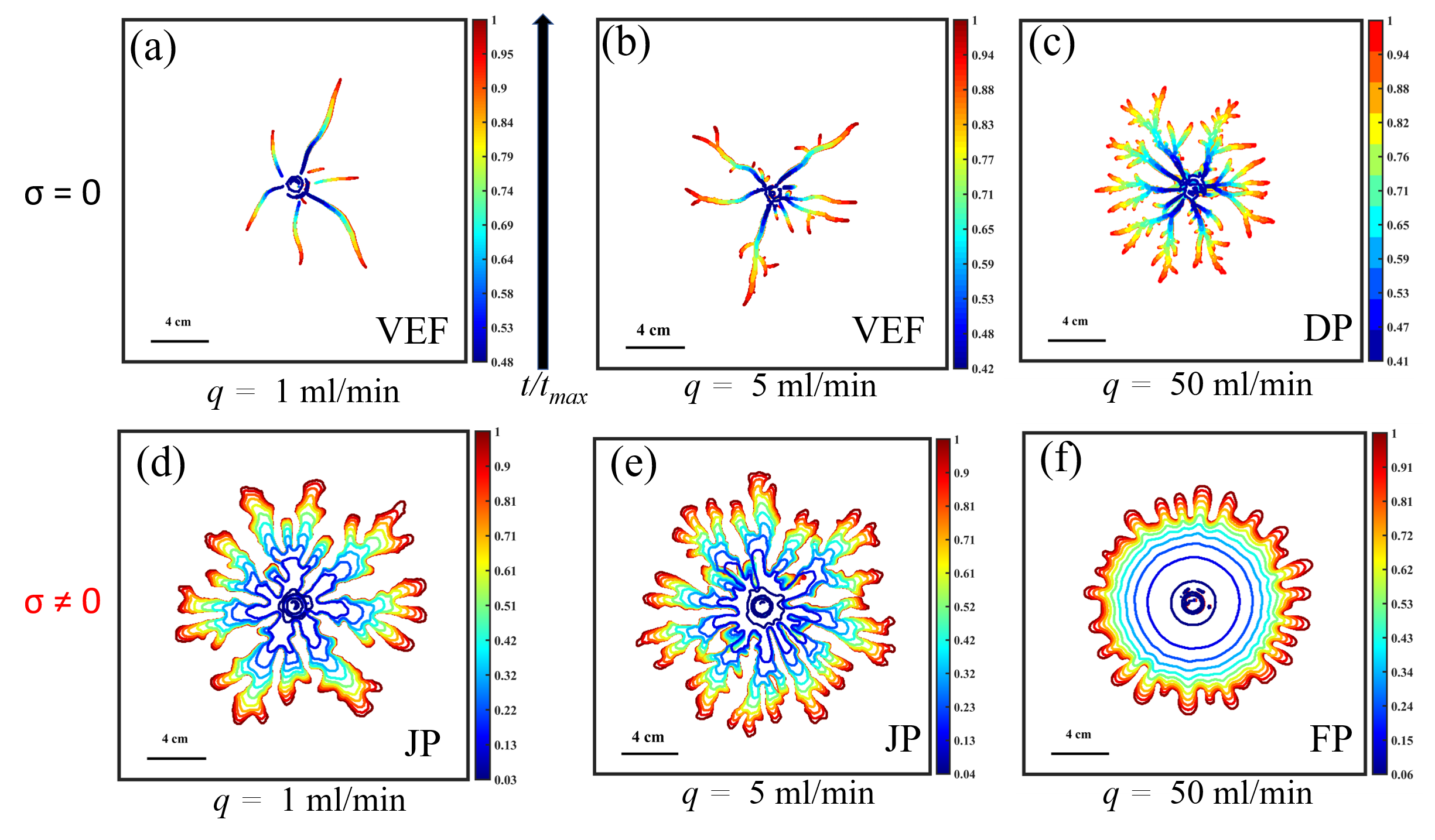}
	\centering
	\caption{\label{fpatternq}\textbf{Temporal evolution of interfacial patterns for different flow rates $q$ of the displacing Newtonian fluid.} Miscible (interfacial tension $\sigma$ = 0) displacement of 3.25 \% w/v aqueous Laponite suspensions ($t_w$ = 24 h) by water at a different flow rates : \textbf{(a,b)} Viscoelastic fractures (VEF) and \textbf{(c)} dendritic pattern (DP). The colourbars map the normalised times $t/t_{max}$, where $t_{max}$ is the total experimental time. Immiscible ($\sigma \neq $ 0) displacement of an aqueous Laponite suspension ($t_w$ = 24 h) of concentration 3.25\% w/v by mineral oil under similar conditions: \textbf{(d,e)} jagged pattern (JP) and \textbf{(f)} flower pattern (FP).}
\end{figure}

Supplementary Figs.~\ref{fpatternq}(a-c) show the temporal evolution of interfacial patterns that form when an aqueous Laponite suspension of fixed age $t_w = 24$h is displaced by water at various flow rates $q$. For miscible displacement experiments at low flow rates, the pattern morphologies lie in viscoelastic fracturing (VEF) regime (Supplementary Figs.~\ref{fpatternq}(a-b)). As the flow rate is increased, pattern growth is dominated by multiple tip-splitting events, and dendritic patterns (DP, Supplementary Fig.~\ref{fpatternq}(c)) are seen to evolve resembling those seen in Fig.~2(b). As in these experiments, we note that even for immiscible displacements, increasing injection flow rates of the displacing fluid results in the same pattern morphologies (Supplementary Figs.~\ref{fpatternq}(d-f)) that are observed by displacing Laponite suspensions of lower $t_w$ at a fixed flow rate (Figs.~3(a,c) of main paper). 

\begin{figure}[!t]
	\includegraphics[width=5.5in]{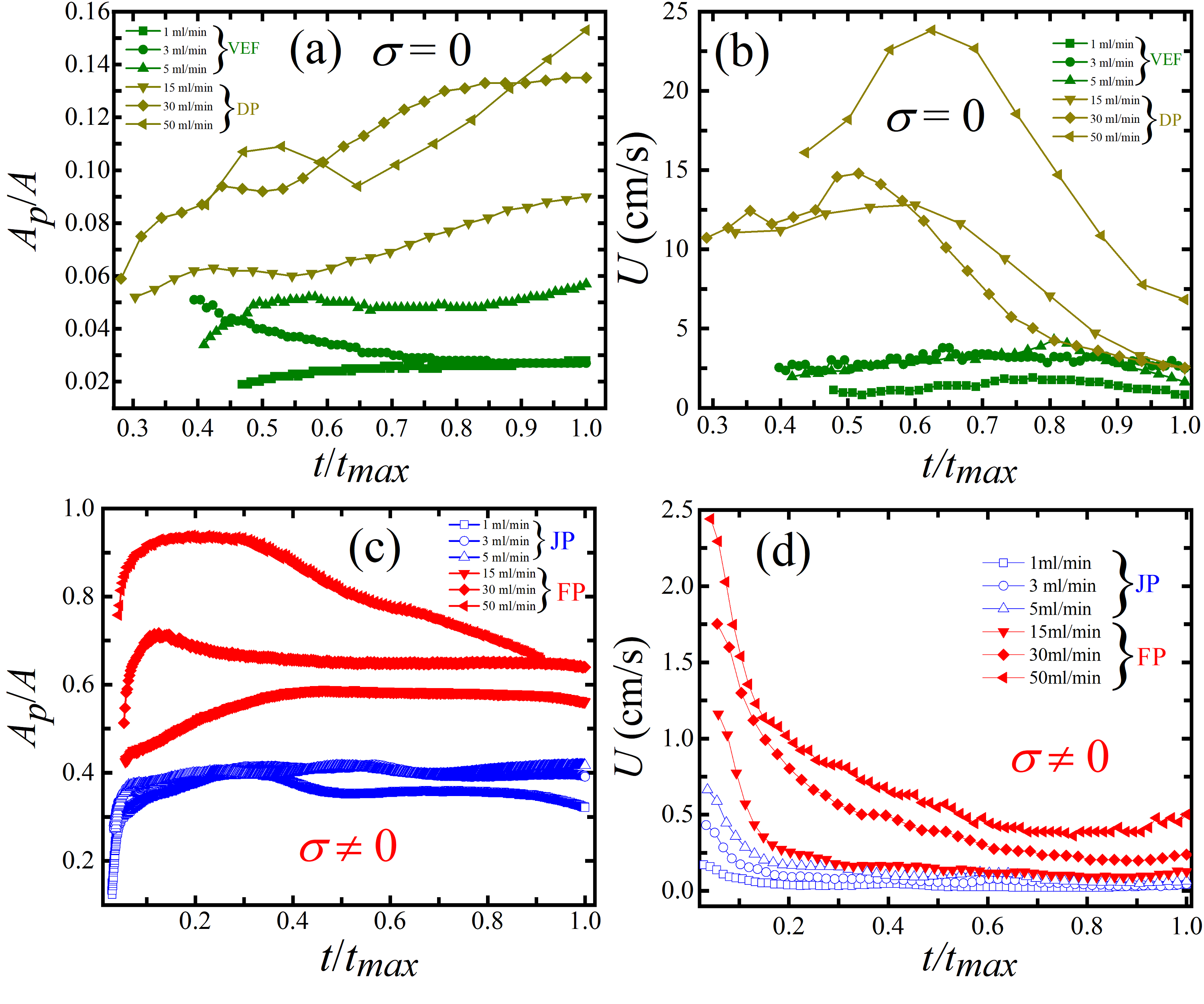}
	\centering
	\caption{\label{ApAq}\textbf{Areal ratios and finger-tip velocities for miscible (interfacial tension $\boldsymbol{\sigma}$ = 0) and immiscible ($\boldsymbol{\sigma \neq}$ 0) patterns with flow rate of the displacing fluid, $q$, as a control parameter.} \textbf{(a)} $A_p/A$ as a function of pattern evolution time, normalised with total experimental time $t_{max}$, for miscible (interfacial tension $\sigma$ = 0) patterns for different $q$ of the displacing fluid. \textbf{(b)} Instantaneous velocity $U$ of the finger-tip of miscible ($\sigma$ = 0) interfacial pattern is plotted against normalised time $t/t_{max}$ for different $q$. \textbf{(c)} $A_p/A$ as a function of normalised time $t/t_{max}$ for immiscible (interfacial tension $\sigma$ = 0) displacement patterns for different $q$. \textbf{(d)} Instantaneous velocity $U$ vs. time $t/t_{max}$ for immiscible ($\sigma \neq $ 0) displacements.}
\end{figure}

The pattern growth is quantified in terms of the pattern areal ratio $A_p/A$ and longest finger-tip velocity $U$. The variations in $A_p/A$ (Supplementary Fig.~\ref{ApAq}(a)) of miscible displacement patterns are similar to the earlier experiments at different suspension aging times, $t_w$ (Supplementary Fig.~\ref{misctw}(a)). The simple inverse correlation between $t_w$ of the displaced suspension and $q$ of the displacing fluid is further confirmed by comparing the similar variations in the instantaneous finger-tip velocities, $U$, in Supplementary Fig.~\ref{ApAq}(b) and~\ref{misctw}(b). The non-monotonic variation in the finger-tip velocity of DP reflects the dominance of shear-thinning effects over elastic ones at high flow rates. We also note that the immiscible jagged patterns (JP) and flower patterns (FP) obtained by increasing $q$ show the same variations in $A_p/A$ (Supplementary Fig.~\ref{ApAq}(c)) and $U$ (Supplementary Fig.~\ref{ApAq}(d)) as seen when aqueous Laponite suspensions of lower $t_w$ were displaced by oil (Supplementary Figs.~\ref{misctw}(c-d)). The waviness in $U$ of the JP persists even for the experiments with varying flow rates (Supplementary Fig.~\ref{ApAq}(d)). 	
	
\end{document}